*CORRESPONDENCE
Urko Peña Alonso
✉ urko.pena@ehu.eus

RECEIVED 05 June 2026
REVISED 21 September 2026
ACCEPTED 30 September 2026
PUBLISHED 21 October 2026

CITATION
Peña Alonso U, Meso Ayerdi K, Peña Fernández S and Morales i Gras J (2026) Cutting out the middleman? Elon Musk and the rhetoric of citizen journalism in platform-mediated communication. Front. Commun. 11:1901417. doi: 10.3389/fcomm.2026.1901417

# Cutting out the middleman? Elon Musk and the rhetoric of citizen journalism in platform-mediated communication

Urko Peña Alonso*, Koldobika Meso Ayerdi, Simón Peña Fernández and Jordi Morales i Gras

Gureiker Research Group (IT1496-22), Department of Journalism, University of the Basque Country, Bilbao, Spain

This article examines how Elon Musk constructed discourse about journalism and the media on X between the 2024 United States presidential election and Donald Trump's presidential inauguration. It investigates how references to journalism, media organisations, and platform-based communication were articulated within Musk's posts, and how users responded to these narratives during a period of intense political polarisation. Particular attention is given to the symbolic appeal to citizen journalism and to the broader question of how informational authority is negotiated in platform-mediated public communication. The study combines computational classification, qualitative thematic coding, and inferential statistical analysis. A corpus of 2,906 posts published by Musk between 5 November 2024 and 20 January 2025 was compiled. From this corpus, 500 posts referring to journalism, media organisations, platforms, or media-related discourse were identified through AI-assisted filtering and subsequent manual review. In addition, 42,066 user responses were analysed from a stratified sample drawn from 3,233,112 interactions. The analysis examined patterns of agreement and disagreement with Musk's stance and their relationship with thematic, gender, and professional variables inferred from publicly available profile information. The findings show that Musk's discourse is organised around three recurrent dynamics: criticism of institutional journalism, the association of media organisations with political and ideological agendas, and the presentation of X as an alternative communicative space. Journalism is predominantly framed through accusations of bias, disinformation, propaganda, and electoral manipulation, whereas X is more often linked to participation, innovation, and direct communication. User responses largely converged with Musk's position, although disagreement remained visible. While thematic orientation and professional background showed statistically significant associations with agreement patterns, effect sizes were limited.

# 1 Introduction

The day after Donald Trump won the US presidential election for a second time, Elon Musk addressed all users of X with the statement: "You are the media now" (@elonmusk, 6 November 2024). With these words, his message appeared, at least on the surface, to evoke early conceptualisations of citizen journalism, whereby audiences, empowered by accessible digital tools, could participate in news production and circulation alongside traditional media institutions (Bowman and Willis, 2003; Bruns, 2021; Gilmor, 2004).

In these early understandings, social media were not merely a technological innovation but a cultural phenomenon that blurred the boundaries between news producers and consumers (Hermida, 2012; Lewis, 2012). Users could produce informational content and assume functions traditionally reserved for journalists (Deuze, 2006). In this sense, participation was understood as a form of power redistribution (Carpentier, 2016), rooted in the involvement of non-professional actors in news production and associated with ideas of authenticity and democratisation (Peña-Fernández et al., 2024).

Digital communication technologies have upended the business models and work practices of professional journalists (Koliska et al., 2025). Digital media have expanded the number of voices participating in the public sphere, weakening the privileged position journalists once held when they controlled the channels of dissemination. As a result, journalists no longer monopolise the production of meaning around current affairs. Instead, they compete and coexist with other social and institutional actors that also produce and distribute information (Belair-Gagnon et al., 2019; Carlson et al., 2021).

Over recent decades, the media have lost part of their authority as public trust has progressively declined (See, 2025), while news consumption has shifted towards digital platforms, which have become central channels for accessing information and forming opinion. In this hybrid media system (Chadwick, 2013) the words and actions of public figures shape public attention at unprecedented speed and scale (Happer and Philo, 2013) and a paradox emerges: the same processes that can encourage civic engagement and campaign participation might also erode institutional legitimacy, foster distrust and partisan divergence (Shah et al., 2017).

This crisis of journalism —or at least of traditional media— cannot be explained solely by internal failures, but also by external attacks that erode its legitimacy as an impartial arbiter of facts (Koliska et al., 2020; Ojala, 2021; Panievsky, 2022). Moreover, this exposure to elite discourse about fake news leads to lower levels of trust in media and less accurate identification of real news (Van Duyn and Collier, 2019).

Accusations of bias have weakened that function, and they have done so in a context in which journalists, as a matter of professional norm, have tended to avoid responding directly to such attacks. On the one hand, journalists have tended to refrain from direct rebuttal, so that while messages discrediting the media abound, messages explaining or defending journalistic work remain scarce (Farkas, 2023). On the other hand, studies of anti-media rhetoric show that attacks by political elites reduce trust in the media, whereas news content itself has only a limited effect on that perception (Watts et al., 1999). In other words, if criticism is constant and defences are rare, trust is likely to erode (Pingree et al., 2018). Journalism, then, is a discursive institution, and its authority is not guaranteed. It is constructed and contested in the public sphere (Koliska et al., 2020).

The rise of populism has been accompanied by growing criticism of the media, both from political leaders and from their supporters (Egelhofer et al., 2021; Marques, 2024). In this discourse, the main catalyst is anti-elitism: media perceived as part of the establishment are evaluated negatively, while those that fit more closely with the cultural framework of populism may be regarded as more trustworthy (Fawzi, 2019). This logic explains why apparently contradictory attitudes can coexist, with distrust towards "the media" in general alongside trust in particular formats or types of content (Hallin et al., 2025).

The construction of the press as an enemy is neither a purely rhetorical nor a conjunctural phenomenon. It is a mechanism with structural implications that alters the institutional balance of democracy by weakening the main actor responsible for scrutinising power and sustaining the public sphere of information. The process of "enemy construction" is not simply an intensification of the historical conflict between politicians and the press, but an operation that redefines the press's status from democratic actor to external, illegitimate and potentially dangerous threat (Jones and Sun, 2017).

Within the digital public sphere, digital press criticism has emerged from a transformed media environment in which multiple actors compete to define truth and journalistic authority remains under constant contestation (Carlson et al., 2020). Criticism of the media is not new. What has changed are the channels through which it is amplified and made more persistent. These platforms also facilitate the formation of networks of actors, including alternative media, users and bots, that reinforce and disseminate such messages.

The open confrontation between political power and journalism, intensified during Donald Trump's presidency, forms part of a discourse that is not new (Egelhofer et al., 2022) but has acquired a different scale and visibility in the contemporary digital environment (Whipple and Shermak, 2020). Shortly after assuming office in January 2017, Trump accused the press of being an "enemy of the American people" (Kalb, 2018).

Trump does not use social media to bypass the media, but to influence the media system as a whole (Meeks, 2020). The end result is a process of selective legitimation. Some media outlets are presented as valid sources —mainly conservative ones— while others are defined as illegitimate actors, or even as enemies. This operation is not merely rhetorical. It has potential effects on public trust and on the media's capacity to fulfil their democratic function. Populism, understood as a strategy of mobilisation, therefore tends to weaken one of the central mechanisms of democratic accountability: the press (Kenny, 2020). In doing so, it seeks to erode the institutional myth that sustains its authority (Koliska et al., 2020).

However, as See (2025) argues, Elon Musk's appeal to citizen journalism was not based on a clear definition of what journalism is, nor on any verifiable transformation of its practices. Rather, it was a discursive device that deployed a term with strong symbolic value, associated with authenticity, proximity and the absence of filters, in order to criticise traditional journalistic mediation and propose an alternative centred on the use of his platform.

Elon Musk thus emerges as an opinion leader who does not belong to the traditional political field (Pallathadka and Pallathadka, 2023), but who combines technological, economic and cultural influence. Musk acts as a globally influential person, with political ambitions and a clear business orientation (Magalhães et al., 2026). His ownership of the dominant social media platform, X (formerly Twitter), enabled him to shape public debates in real-time (Stambach, 2025). This gives him the capacity to shape the public agenda and collective attention through his ability to attract attention and reorganise the flow of information without occupying formal positions within the political system (Gao et al., 2025). This type of intervention does not replace traditional actors, but it introduces an additional layer into the formation of public opinion, where attention becomes a central resource in political competition. This power to direct information flows has been particularly evident during major events such as elections, as was seen during the 2024 US presidential election.

This study examines how Elon Musk constructs discourse about the media on X during the period between the 2024 US presidential election and Donald Trump's presidential inauguration. The analysis focuses on the semantic organisation of Musk's messages, the interaction patterns they generate among users, and the extent to which these responses vary according to users' sociodemographic characteristics.

The case is relevant for at least three reasons. First, Musk occupies an unusual position at the intersection of platform ownership, political communication, and algorithmically amplified visibility. Second, his public interventions increasingly blur the boundaries between media criticism, partisan communication, and platform governance. Third, his discourse about journalism does not operate merely as commentary on the media system. It actively participates in the symbolic redefinition of informational authority by contrasting institutional journalism with decentralised, platform-based forms of communication.

Rather than treating criticism of the media as episodic or reactive, Musk's activity on X suggests a more stable discursive structure organised around distrust towards legacy journalism, accusations of political manipulation, and the promotion of platform-mediated disintermediation. Within this framework, X is not presented simply as a technological infrastructure but as a normative alternative to institutional media.

Existing research on populist communication, anti-media rhetoric, and platform politics has extensively documented the delegitimisation of journalism by political actors. Less attention has been paid to cases in which these dynamics are articulated by platform owners whose communicative power derives simultaneously from political visibility, economic influence, and infrastructural control over digital public space. Musk represents a particularly significant case because he combines these dimensions while directly intervening in public debates through highly personalised communication practices.

In light of this, the study is guided by the following research questions:

RQ1. How are Elon Musk's messages distributed according to the type of media entity involved, and which actors occupy the most prominent positions?

RQ2. How are the media discursively constructed within Elon Musk's narrative, and which attributes are associated with them?

RQ3. Do distinct semantic strategies emerge in references to news media organisations and to the social platform X?

RQ4. How can users' interactions with Elon Musk be characterised, and which patterns of convergence or divergence can be identified?

RQ5. Do users' responses to Elon Musk's position display differentiated patterns according to sociodemographic variables?

# 2 Materials and methods

This study combines computational classification, qualitative thematic coding, and inferential statistical analysis to examine how Elon Musk constructs discourse about the media on X and how users respond to those narratives. The methodological design was organised in four stages: corpus construction, thematic classification of Musk's posts, collection and analysis of user responses, and statistical association analysis.

Posts published from 5 November 2024, the date of the US presidential election, through 20 January 2025, when Donald Trump was inaugurated as President of the United States, were collected for analysis. This period captures the presidential transition from the election outcome to the beginning of the new administration. A phase marked by high levels of political polarisation and intense public debate.

The corpus included both Musk's original posts and the posts he quoted or reposted. This decision was analytically necessary because Musk frequently communicates through short comments attached to third-party publications rather than through standalone messages. In these cases, meaning is often produced relationally through endorsement, irony, amplification, or rejection of previously circulating content. Limiting the analysis to Musk's original text would therefore exclude a substantial part of his communicative practice.

In total, 2,906 posts were compiled.

## 2.1 Identification of posts referring to news media

The first stage consisted of identifying posts containing references to journalism, news organisations and platforms. The screening process considered four dimensions simultaneously: the textual content of Musk's posts, the content of quoted publications, the identity of referenced actors, and the semantic framing surrounding those references. Posts were collected from X using Apify through web-scraping techniques rather than through direct API access.

To assist the preliminary filtering process OpenAI's, GPT-4o mini model was used through API access. The model was employed exclusively as a support tool for large-scale semantic identification and preliminary categorisation. It was not used as an autonomous classification mechanism.

AI-assisted filtering was adopted solely to manage the scale of the 2,906-post corpus. Because nuanced social media features—

such as irony, slang, and implicit references require human interpretation, researchers maintained full control over the classification pipeline. Beyond manually validating every post flagged as relevant by the model to establish the final corpus ($n = 500$, 17.2%), researchers also conducted a manual quality audit on a random 10% sample of the discarded posts, confirming that no systematic false negatives were introduced during the automated triage phase.

A structured prompt was designed to identify whether each post included explicit or implicit references to media organisations, journalism, platforms, or media-related discourse. The prompt was iteratively refined during pilot testing to reduce ambiguous classifications.

All outputs generated by the model were subsequently reviewed manually by the researchers. Cases involving semantic ambiguity, humour, sarcasm, or unclear contextual references were re-evaluated independently before final classification. Through this combined procedure, 500 posts were identified as relevant to the study.

## 2.2 Thematic classification of Elon Musk's posts

Once the relevant posts had been identified, the material underwent thematic coding aimed at detecting recurrent discursive patterns within Musk's communication about the media. The analysis identified eight major thematic dimensions: criticism of media bias and lack of credibility; accusations of disinformation and propaganda; defence of freedom of expression and opposition to censorship; promotion of X as a direct communication channel; humour and meme-based delegitimisation; claims of political or electoral manipulation; technological transformation and artificial intelligence; and calls for media reform and accountability.

The coding strategy followed a multilabel approach, allowing individual posts to be assigned to multiple categories when several themes coexisted within the same publication. This decision was necessary because many posts simultaneously combined political criticism, anti-media rhetoric, humour, and platform promotion. The objective was not merely to identify isolated topics, but to capture the discursive articulation through which these elements appeared interconnected within Musk's communication practices.

Two researchers independently coded the corpus, which was divided equally between them, with each researcher coding half of the posts. To assess inter-coder reliability, an overlapping set of 50 texts was embedded as a blind control subset within both researchers' coding assignments. Agreement between the two coders was substantial, yielding a Cohen's kappa of $\kappa = 0.72$ and an observed agreement of 0.83. Discrepancies within this control subset, as well as disagreements arising during the coding process, were subsequently discussed and resolved through consensus review. Rather than relying exclusively on the presence of keywords or explicit references, the final coding scheme prioritised semantic function and contextual meaning. This approach made it possible to classify posts containing irony, implicit associations, memes, or indirect forms of delegitimisation that would have been difficult to identify through purely lexical procedures.

## 2.3 Collection of responses to Elon Musk's posts

User responses to Musk's posts were collected using the Apify platform. In total, 3,233,112 interactions associated with the selected posts were identified.

Given the scale of the dataset and the operational limitations associated with API extraction and processing, a probabilistic sampling strategy was implemented. The final sample ($n = 42{,}066$) was generated through multidimensional stratification to preserve variability across users and temporal distribution. The procedure aimed to reduce distortions associated with highly active accounts, coordinated participation, or temporally concentrated interaction peaks (Vicente, 2023). The resulting sample preserved diversity across interaction dynamics while remaining computationally manageable for individual-level semantic analysis.

## 2.4 Analysis and classification of user responses

The sampled responses were analysed individually using OpenAI's GPT-4o mini model through API access. The analytical prompt incorporated four elements simultaneously: Musk's original publication, the quoted post where applicable, the user response, and publicly available profile metadata.

The analysis focused on two principal dimensions. First, the model classified the degree of agreement or disagreement with Musk's position. Responses were categorised as agreement, disagreement, or non-identifiable positioning when no clear evaluative orientation could be inferred.

Second, the model inferred users' gender and professional category when sufficient information was available in profile descriptions. These variables were treated as approximate indicators derived from self-presentational information rather than as verified demographic attributes.

Because automated inference involving social media profiles inevitably introduces uncertainty, particular caution was applied during interpretation. The analysis therefore does not claim to identify users' actual identities or occupations, but rather the publicly projected social signals available within platform interactions.

## 2.5 Strengths and limitations of the AI-assisted methodology

The methodological incorporation of OpenAI's GPT-4o mini across two distinct stages—first, the preliminary identification of posts on X referencing journalism, media organizations, and digital platforms; and second, the classification of user replies—provided critical operational efficiencies. In both phases, automated semantic assistance resolved substantial scale constraints. This dual-stage computational support facilitated a systematic, high-throughput triage of social media discourse that would have been logistically unfeasible through purely manual coding.

Nevertheless, this approach presented methodological limitations specific to each analytical stage. During initial identification, the model could occasionally misinterpret implicit media criticism, platform-specific language, evolving digital jargon, and ambiguous references. In reply analysis, it was particularly vulnerable to irony, cynicism, memes, elliptical phrasing, and sarcasm when comments were considered outside their broader conversational context. To mitigate these risks, the model was used solely as a heuristic aid, while researchers manually verified and independently re-evaluated all ambiguous, borderline, and context-dependent outputs.

## 2.6 Statistical analysis

The inferential stage focused exclusively on responses classified as agreement or disagreement. Interactions categorised as non-identifiable positioning were excluded in order to prioritise analytically interpretable responses.

Contingency tables and Chi-square tests were used to examine associations between variables. Cramér's V coefficient was calculated to estimate effect size and avoid overinterpreting statistically significant associations with limited substantive magnitude.

Particular attention was paid to the distinction between statistical significance and explanatory strength. Given the large sample size, relatively small differences could produce statistically significant results. Consequently, the interpretation of findings prioritised effect size and distributional tendencies over significance values alone.

# 3 Results

## 3.1 Discursive structure of Elon Musk's references to the media

To address RQ1 and RQ2, the analysis first examines which media actors appear in Musk's posts and the thematic frames through which journalism is constructed.

The analysis of the corpus ($n = 500$) reveals that Musk's discourse about the media operates simultaneously on two levels. The first is abstract and systemic, built around generic categories such as "media", "legacy media", "mainstream media", and "press", which together account for more than one third of all references identified in the corpus. The second level is more concrete and illustrative, centred on explicit references to identifiable organisations and platforms. In this second group, television networks and large digital media outlets occupy a particularly visible position, alongside X itself, which appears more frequently than other social platforms such as Facebook, LinkedIn, or YouTube (Table 1). The table also includes The Babylon Bee, a conservative satirical outlet often described as a conservative counterpart to The Onion. Although classified as a Digital Media Outlet, this categorisation refers to its digital-native format and does not imply equivalence with conventional digital news organisations.

TABLE 1 Media typology and frequency of the 10 most mentioned actors.

| Specific/general reference | Category | Frequency |
|---|---|---|
| Media | Generic | 96 |
| Legacy media | Generic | 60 |
| X | Social media platform | 21 |
| GBNews | Television | 14 |
| CNN | Television | 12 |
| The Babylon Bee | Digital media outlet | 12 |
| Mainstream media | Generic | 12 |
| The New York Times | Press/newspaper | 11 |
| Press | Generic | 11 |
| Fox News | Television | 10 |

Source: Authors' own elaboration.

This distinction is not merely descriptive. Generic references allow Musk to construct journalism as a homogeneous institutional actor associated with political coordination and informational manipulation, whereas references to specific organisations function as concrete examples that validate broader claims about bias, propaganda, or ideological alignment. The recurrence of terms like "traditional media" is especially significant because it encapsulates the critique of journalism and the traditional conception of mediation, in contrast to the seemingly disintermediated communication of social networks.

The distribution of actors also reveals a strong geographical concentration. Most identifiable references correspond to US media organisations ($n = 153$), followed at a considerable distance by British outlets ($n = 42$). By contrast, continental European organisations appear only occasionally. This pattern situates Musk's anti-media discourse primarily within the communicative conflicts of the Anglo-American political sphere, particularly the United States during the presidential election cycle.

At the thematic level, criticism of media credibility and accusations of political bias clearly dominate the corpus, appearing in 59.2% ($n = 296$) of the analysed posts (Figure 1). Across these messages, journalism is repeatedly framed as selective, dishonest, partisan, or structurally incapable of neutrality. These criticisms rarely target isolated professional failures. Instead, they portray bias as an intrinsic characteristic of institutional media systems.

Accusations of disinformation and propaganda constitute the second most visible thematic axis, appearing in approximately one third of the corpus (35%) ($n = 175$). Here, journalism is represented not simply as inaccurate, but as intentionally deceptive. References to lying, concealment, coordinated messaging, and manipulation recur systematically, often connected to electoral politics and ideological conflict.

Themes related to freedom of expression and censorship appear with very similar frequency. In these posts, media organisations are portrayed as actors that regulate visibility, suppress dissenting positions, or protect dominant political narratives. Criticism of journalism therefore extends beyond questions of accuracy or credibility and becomes linked to broader disputes concerning informational control and legitimacy within the public sphere.

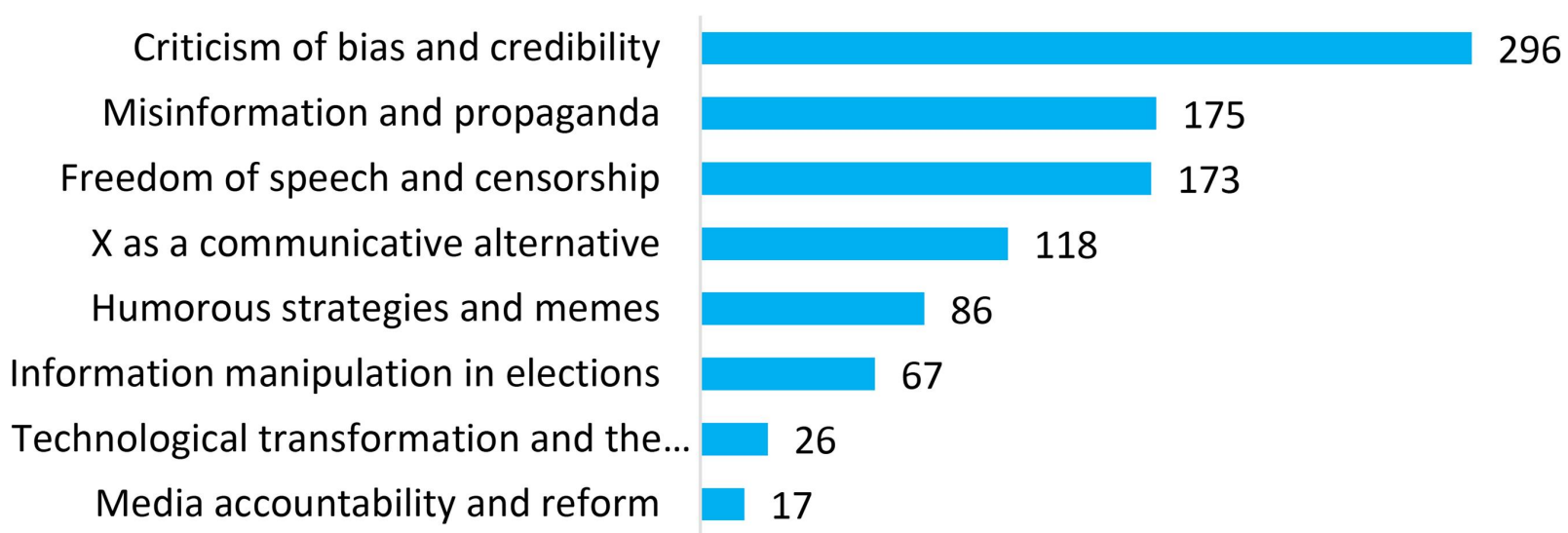

FIGURE 1
Main discursive themes identified in the analysed posts. Source: Authors' own elaboration.

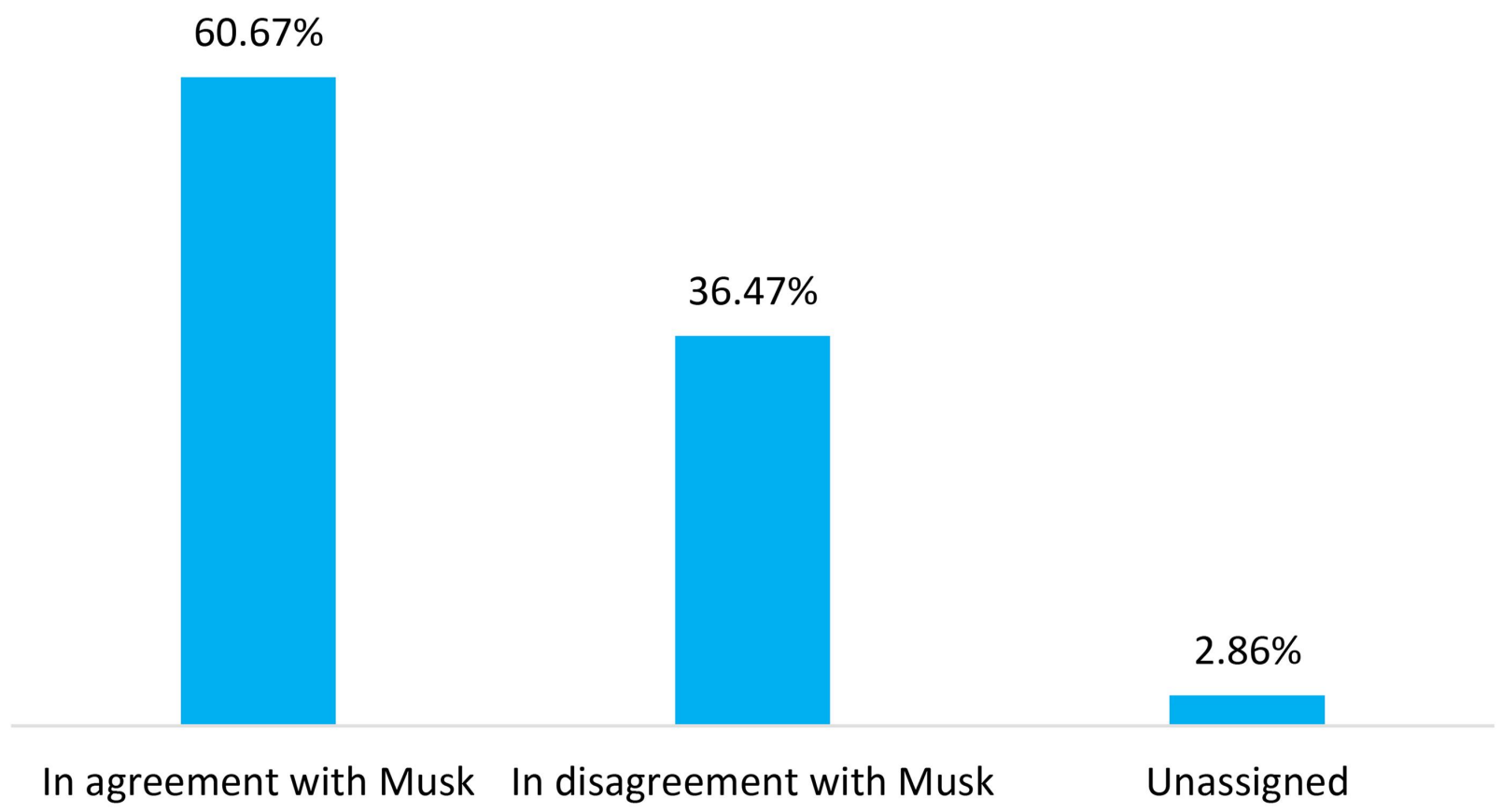

FIGURE 2
Distribution of responses according to users' positioning towards Musk's stance. Source: Authors' own elaboration.

Another important dimension concerns the representation of X as a communicative alternative to institutional journalism (23.6%) ($n = 118$). References associated with disintermediation, direct participation, and citizen-led information flows appear in nearly one quarter of the analysed posts. Expressions such as "you are the media now" illustrate how Musk relocates informational authority from professional journalism to networked participation and platform-based interaction.

Humour and memes also occupy a relevant position within the corpus (17.2%) ($n = 86$). Although quantitatively less central than anti-media criticism or disinformation claims, they appear in almost one fifth of the posts analysed. Their function is not secondary. Sarcasm, ironic renaming, and meme-based ridicule operate as mechanisms of symbolic delegitimisation directed at media organisations and journalists.

Other themes appear with lower frequency in the corpus. Claims concerning media manipulation in electoral contexts account for 13.4% ($n = 67$) of the analysed posts, particularly in relation to the United States, the United Kingdom, and Germany. Discussions surrounding technological transformation and artificial intelligence in the future of communication represent 5.2% ($n = 26$) of the corpus, while calls for reform of the media system appear in 3.4% ($n = 17$) of the analysed posts.

When references to "the media" are examined as an abstract category (see Table 2), several differences emerge in comparison with the overall corpus. The most prominent thematic axis becomes allegations of media manipulation in electoral processes, accounting for 33% of the references to media in general. In these messages, media organisations are framed less as informational institutions than as partisan political actors. They are represented as protectors of the establishment,

TABLE 2 Thematic axis, frequency, and percentage for references to media in general.

| Thematic axis | Frequency | Percentage |
|---|---|---|
| Media manipulation in elections | 62 | 33% |
| Disinformation and propaganda | 46 | 24.5% |
| Criticism of bias and credibility | 45 | 23.9% |
| X as a communicative alternative | 42 | 22.3% |
| Humour and meme strategies | 16 | 8.5% |
| Media accountability and reform | 11 | 5.9% |
| Freedom of expression and censorship | 10 | 5.3% |
| Technological and media transformation | 3 | 1.6% |

Source: Authors' own elaboration.

amplifiers of ideological agendas, agents intervening in electoral processes, or actors aligned with particular political parties.

Within this narrative, several recurrent semantic associations appear across the corpus, including references linking the media to "Democrats–Biden–Obama", "Labour–Starmer", "Soros", "AfD", "immigration", and the "political establishment". These associations situate criticism of journalism within broader political conflicts and reinforce the representation of media organisations as active participants in ideological struggle rather than neutral intermediaries.

The following three thematic categories occupy relatively similar positions: disinformation and propaganda (24.5%), criticism of bias and credibility (23.9%), and X as a communicative alternative (22.3%). Together, these themes reinforce a broader discursive structure in which journalism is associated with political manipulation, disinformation, and lack of credibility, while X is positioned as an alternative communicative space.

## 3.2 Journalism and X as competing communicative models

RQ3 examines whether distinct semantic strategies emerge in references to news media and X. The comparison reveals a marked asymmetry between the two.

Although direct references to X represent a relatively limited proportion of the corpus, the platform occupies a central position within Musk's broader discourse about the media. Beyond the posts explicitly mentioning X, the platform also appears indirectly throughout the corpus as the implicit communicative alternative to institutional journalism. This produces a clear asymmetry between the semantic construction of journalism and that of the platform itself.

This cross-sectional analysis (see Table 3) reveals substantial differences between the discursive construction of journalism and that of the social platform X. References to journalism are predominantly organised through delegitimising frames associated with manipulation, ideological bias, propaganda, and political coordination. By contrast, references to X are systematically linked to notions of openness, participation, technological innovation, and communicative freedom. The contrast is therefore not only technological but normative: journalism and the platform are presented as competing models of informational authority.

TABLE 3 Key discursive differences between the platform X and the media.

| Aspect | Media | Social platform X |
|---|---|---|
| Dominant discourse | Delegitimisation: manipulation, propaganda, bias, deception | Legitimation: disintermediation, innovation, and freedom |
| Technology and innovation | 7.6% | 40.2% |
| Political criticism | 33.2% | 24.5% |
| Disinformation | 28.5% | 19.5% |
| Freedom of expression | 12.3% | 2.4% |
| Humour and satire | 8.6% | 7.3% |
| Reform and accountability | 9.8% | 6.1% |

Source: Authors' own elaboration.

The clearest divergence appears in relation to technological and media transformation. While references linking the media to innovation remain limited (7.6%), this dimension occupies a much more central position in messages concerning X (40.2%), where the platform is repeatedly presented as a space of technological development and communicative renewal.

Political criticism presents a more balanced distribution, accounting for 24.5% of references to X and 33.2% of those directed at the media. In both cases, these criticisms are concentrated primarily around references to the US Democratic Party and its political representatives.

Expressions related to disinformation also differ substantially in their semantic orientation. References to X frequently portray the platform as a corrective mechanism or antidote to disinformation (19.5%), embedded within a positive narrative centred on transparency and solutions. By contrast, media organisations are represented as active participants in the production and circulation of misleading information (28.5%), often portrayed as knowingly disseminating distorted or manipulated narratives.

Themes related to freedom of expression are more strongly associated with references to the media (12.3%), although predominantly within a negative frame linked to censorship, concealment, restrictions on public debate, and forms of social engineering. References to X within this category are less frequent (2.4%) but operate in the opposite direction. Here, both the platform and Elon Musk himself are connected to positive narratives surrounding the defence of free expression.

Humour and satire appear with relatively similar frequency in both cases (8.6% for the media and 7.3% for X), although their discursive function differs substantially. When directed at media organisations, humour operates primarily as a mechanism of delegitimisation through ridicule, ironic renaming, and sarcasm, as reflected in expressions such as "Should be called Disinformation News Network! DNN, not CNN!" or "The New Woke Times instead of The New York Times". In references associated with X, humour instead reinforces narratives of virality and citizen participation, illustrated by comments such as "X as the new centre" or "These are the citizen journalists".

Finally, posts addressing reform and accountability reveal another important asymmetry. References to journalism frequently emphasise the need for correction mechanisms and stronger accountability practices, illustrated by comments such as: "When stories are proven false, corrections must be made promptly and carry as much visibility… as the original piece". By contrast, X is presented as a corrective alternative to what Musk frames as the "interested" mediation of traditional media. Particularly through Community Notes, the platform is represented as an environment capable of collectively monitoring and challenging media narratives, as reflected in statements such as: "Community Notes and rebuttals by users on this platform are the biggest factor pushing the legacy media to be less deceptive".

## 3.3 Patterns of user agreement and disagreement

RQ4 concerns the patterns of convergence and divergence in user responses to Musk's posts. Across the analysed sample, agreement predominates, although disagreement remains substantial.

The analysis of user responses reveals a clear predominance of agreement with Musk's positions (see Figure 2). Of the 42,066 comments included in the sample, 25,520 (60.67%) expressed support for Musk's stance, whereas 15,342 (36.47%) reflected disagreement. Only a small proportion of interactions (2.86%) could not be classified because they did not contain a sufficiently identifiable evaluative position.

User responses largely align with Musk's positions, particularly in posts centred on criticism of journalism and media institutions. Supportive comments clearly outnumber opposing ones, pointing to a strong tendency towards convergence around his framing of the media. Still, the imbalance is not large enough to describe the interaction space as uniform or politically closed.

Disagreement also occupies a visible place within the reply network. More than one third of the analysed responses challenge Musk's positions, showing that his posts do not circulate only within environments of ideological reinforcement. Alongside endorsement, the interaction space also sustains disagreement, confrontation, and direct contestation.

This coexistence of agreement and opposition becomes especially relevant within the political context in which these interactions emerged. The analysed period — between the 2024 US presidential election and Donald Trump's inauguration — concentrated intense public debates surrounding journalism, platform governance, political manipulation, and freedom of expression. User responses therefore appear connected not only to Musk as an individual actor, but also to broader disputes over informational authority and media legitimacy.

The analysis of standardised residuals reveals that agreement and disagreement with Musk are not distributed uniformly across thematic categories (Figure 3). Instead, some themes display clear deviations from expected frequencies, indicating differentiated patterns of user alignment.

The strongest positive association with agreement appears in posts related to political and electoral manipulation and criticism of bias and credibility in traditional media. These categories show a higher-than-expected concentration of supportive responses, suggesting that narratives linking journalism with political intervention or institutional distrust generate stronger convergence around Musk's positions.

By contrast, themes related to X as a direct alternative and debates surrounding technology, artificial intelligence, and the future of communication display the opposite pattern. In these categories, disagreement appears substantially overrepresented, while agreement occurs less frequently than expected. Although support for Musk remains present across the broader corpus, posts centred on technological innovation and platform transformation seem to generate weaker identification dynamics and greater contestation.

Other themes occupy intermediate positions. References to freedom of expression, censorship, disinformation, and propaganda produce only modest deviations from expected distributions, indicating more balanced interaction patterns. Humour and ridicule display a slight tendency towards disagreement, while themes concerning media reform and accountability generate relatively limited differences.

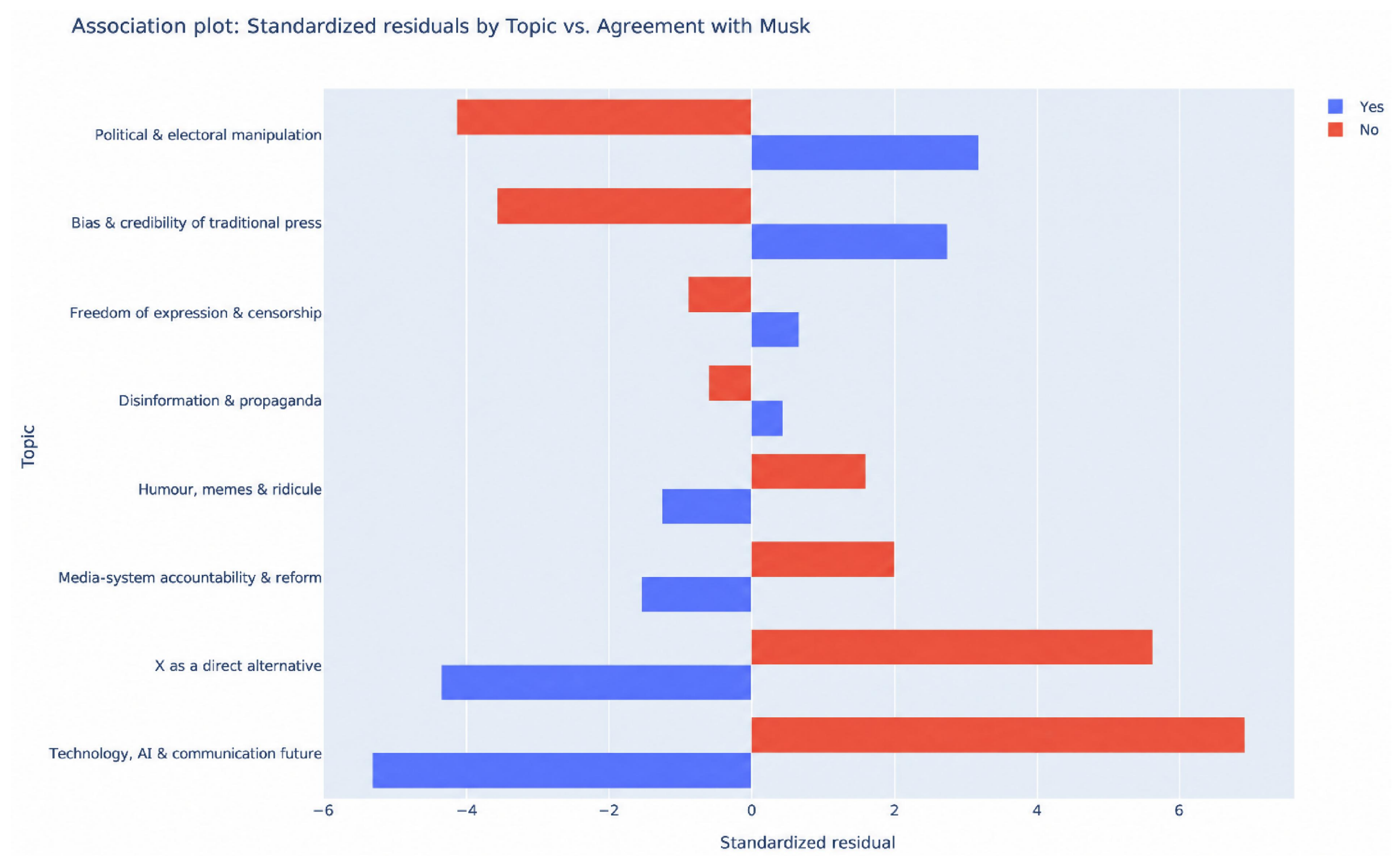

FIGURE 3
Standardised residuals for thematic categories. Source: Authors' own elaboration.

Taken together, the results reveal a relatively coherent relationship between Musk's discursive construction of journalism and the interaction dynamics generated around his posts. The strongest levels of user alignment emerge in messages centred on political manipulation, media bias, and institutional distrust, suggesting that anti-media narratives articulated through conflict and delegitimisation possess greater mobilising capacity than posts focused primarily on technological transformation or platform innovation.

## 3.4 Sociodemographic and thematic variation in user responses

Across the total analyzed sample, users for whom gender inference was possible and were classified as female represented 23.45% ($n = 9,866$), while users classified as male accounted for 16.84% ($n = 7,086$). Profiles assigned to other categories were statistically marginal, while in the remaining 59.68% of accounts ($n = 25,104$), no reliable inference could be established due to insufficient profile information. Consequently, the findings presented here should be interpreted as approximate indicators derived from self-presentational information rather than as verified demographic classifications.

Among the profiles for which gender inference was possible, users classified as female represented 23.45% of the total sample ($n = 9,866$), while users classified as male accounted for 16.84% ($n = 7,086$). Profiles assigned to other categories were statistically marginal.

The inferential analysis reveals no statistically significant association between inferred gender and agreement with Musk's position (Table 4). Although the obtained $p$-value (0.0620) approaches the conventional threshold for statistical significance, the effect size remains extremely limited (Cramér's V = 0.0144). Consequently, the results do not support the existence of a meaningful relationship between gender and users' positioning in favour of or against Musk's discourse.

A similar limitation affects the analysis of professional background (Figure 4). In most cases (81.73%; $n = 34,379$), user profiles did not contain sufficient information to infer occupational affiliation. Nevertheless, the subset of identifiable profiles ($n = 7.043$) reveals several recurrent clusters that help characterise the interaction environment surrounding Musk's posts.

TABLE 4 Association tests between "agreement with Musk" and the analysed variables.

| Variable | Chi-square | $p$-value | Cramér's V |
|---|---|---|---|
| User gender | 3.48 | 0.0620 | 0.0144 |
| Thematic categories | 186.62 | <0.0001 | 0.0485 |
| Professional categories | 91.50 | <0.0001 | 0.1100 |

Source: Authors' own elaboration.

The largest identifiable category corresponds to heterogeneous professional profiles that could not be grouped into more specific sectors, accounting for approximately half of the classifiable cases. Beyond this broad category, the most visible groups include users linked to political and social activism (15.91%), design and content creation (8.33%), information technology professions (7.38%), military and security-related occupations (4.24%), entrepreneurial and management activities (3.39%), and health-related professions (3%).

Professional background also displays a statistically significant association with users' positioning ($p < 0.0001$), with a somewhat higher effect size than that observed for thematic categories (Cramér's V = 0.1100). The analysis of residual distributions indicates greater levels of agreement among users associated with retirement, family-oriented identities, and property-related activities. Lower levels of agreement appear among profiles linked to design and creative industries, scientific and research activities, and political activism or social analysis.

The findings should nevertheless be interpreted cautiously. Professional categories inferred from user profiles do not necessarily correspond to stable occupational positions, and in many cases they reflect aspirational identities, symbolic affiliations, or selective forms of self-description rather than verifiable employment status.

The analysis of standardised residuals also reveals differentiated patterns of agreement and disagreement according to inferred professional background, although the overall effect size remains modest (Figure 5). Several professional categories deviate noticeably from expected distributions, indicating varying levels of alignment with Musk's positions.

The strongest positive association with agreement appears among retired users and profiles centred on family or personal roles. Real estate-related occupations also display higher-than-expected levels of support. Similar, although less pronounced, patterns emerge among users associated with consulting activities, military and security professions, human resources, and agriculture-related sectors. Across these groups, supportive responses appear more frequently than expected under conditions of statistical independence.

By contrast, the strongest deviations towards disagreement are concentrated in a different set of occupational categories. Users associated with art, design, and content creation exhibit the largest overrepresentation of disagreement, closely followed by profiles linked to science and research outside academia, and politics, activism, and socio-political analysis. Smaller but still visible tendencies towards disagreement also appear among students, media and journalism-related profiles, and service occupations.

Between these two poles, several categories — including technology, health, legal professions, and business-related activities — remain closer to expected distributions, displaying comparatively weaker deviations in either direction.

Rather than pointing to sharply segmented communities, these patterns suggest differentiated tendencies in the reception of Musk's discourse. Support appears comparatively stronger among profiles associated with more traditional, managerial, or institutionally structured environments, whereas disagreement concentrates more heavily among users connected to creative sectors, knowledge production, and explicitly political forms of engagement.

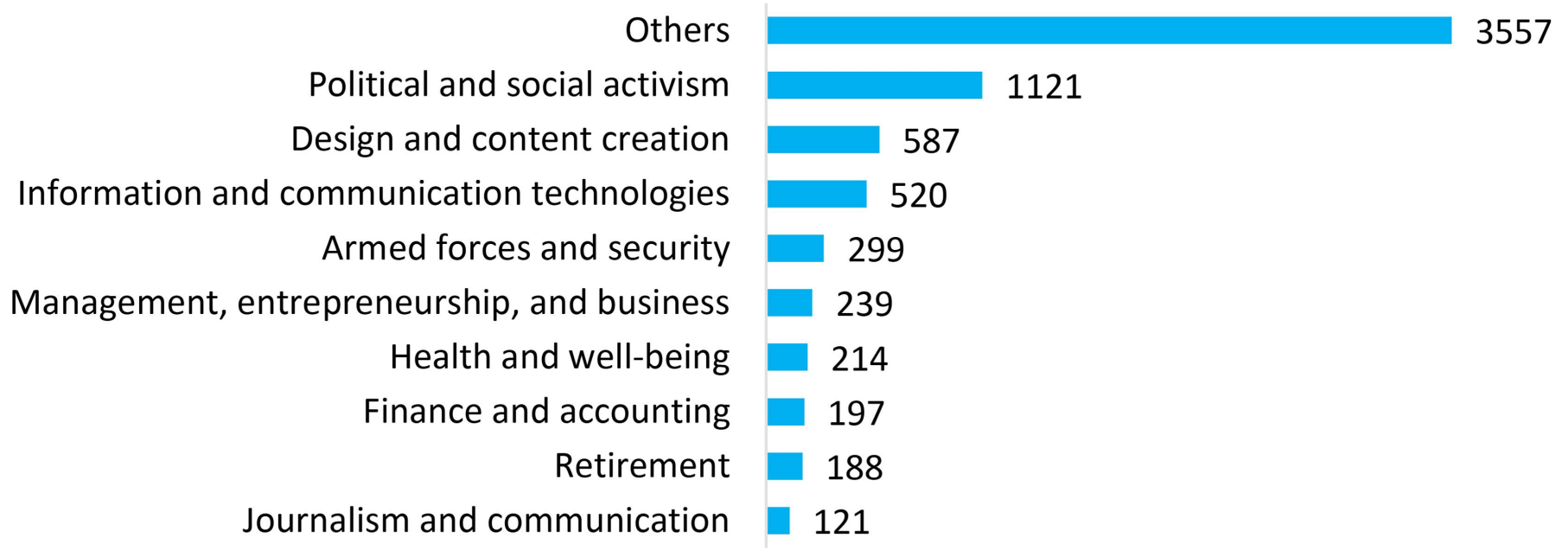

FIGURE 4
Distribution of responses according to professional group. Source: Authors' own elaboration.

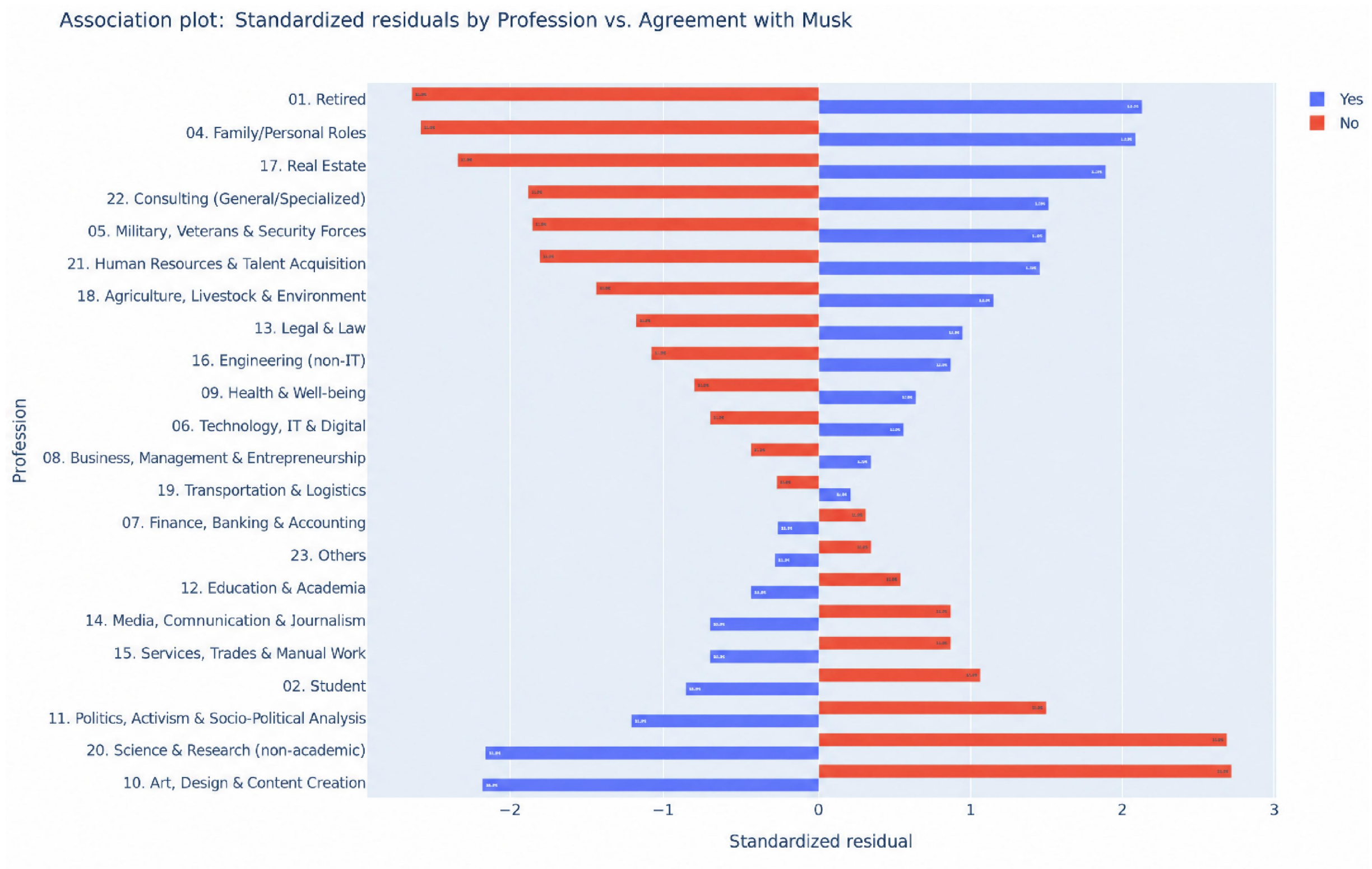

FIGURE 5
Standardised residuals by professional group. Source: Authors' own elaboration.

# 4 Discussion

The findings indicate that Elon Musk's discourse about journalism during the period analysed is organised around a relatively coherent communicative structure rather than around isolated criticism of particular media organisations or journalistic events. Across the corpus, three recurrent dynamics emerge: criticism of institutional journalism, the association of media organisations with political actors and ideological agendas, and the presentation of X as an alternative communicative space (RQ1, RQ2).

The analysis shows that references to journalism are predominantly articulated through delegitimising frames. Generic categories such as legacy media, mainstream media, and media occupy a central position in the corpus and frequently appear linked to accusations of bias, disinformation, political coordination, and electoral intervention. Rather than focusing primarily on specific professional failures or individual news organisations, these references construct journalism as a broader institutional category associated with distrust and political conflict. This pattern is consistent with previous research showing how external attacks on journalism can challenge its institutional authority and legitimacy (Koliska et al., 2020). It also resembles the selective legitimation described by Meeks (2020), in which criticism is directed at "the media" as a general category while particular outlets or sources may retain legitimacy. In Musk's case, however, this discourse is articulated by a platform owner rather than by a political officeholder. This distinction matters because ownership of X places Musk not only within public communication but also in a position to shape the conditions under which content gains visibility and circulates on the platform (Stambach, 2025).

The findings also reveal an important asymmetry in the discursive construction of journalism and X (RQ3). Whereas references to media organisations are mainly associated with manipulation, propaganda, and credibility problems, references to X are more frequently connected to participation, technological innovation, and direct communication. This contrast suggests the coexistence of two differentiated communicative models within Musk's discourse: one associated with institutional mediation and another organised around platform-based interaction.

This distinction becomes visible across several thematic dimensions. References to disinformation frequently portray media organisations as actors involved in the circulation of distorted narratives, while references to X are more commonly associated with corrective dynamics such as participation and user intervention. Similarly, themes related to freedom of expression appear in markedly different ways depending on whether they concern journalism or the platform itself. This asymmetry also qualifies Musk's appeal to citizen journalism. Early accounts associated citizen journalism with the expansion of opportunities for non-professional actors to participate in news production and circulation (Bowman and Willis, 2003). Here, participation is articulated alongside criticism of traditional journalistic mediation: the positive construction of users and X coincides with the negative construction of institutional media. This corresponds with See's (2025) interpretation of Musk's use of citizen journalism as a symbolic device associated with authenticity, proximity, and the absence of traditional journalistic filters, while the present analysis identifies the semantic opposition through which that appeal is articulated.

At the same time, the contrast between institutional media and X does not amount to the disappearance of mediation. As the hybrid media system perspective indicates, established media, digital platforms, and new communicative actors coexist and compete for visibility and influence within the same information environment (Chadwick, 2013). Moreover, platform-based communication involves a different form of mediation. Twitter's algorithmic timeline curates users' exposure to information by selecting and ordering content rather than simply transmitting it directly between producers and audiences (Bandy and Diakopoulos, 2021). Thus, the directness attributed to X in Musk's discourse should not be equated with the absence of mediation. In the discourse analysed here, X is presented as an alternative to journalistic mediation while remaining itself the infrastructure through which communication is organised. This distinction also has an ownership dimension. The platform presented as an alternative to journalistic mediation is controlled by the same actor who promotes it as such. This concentration of platform ownership matters because ownership can provide individual platform proprietors with direct influence over content governance (Leerssen, 2025).

User responses largely converged with Musk's positions (RQ4). Approximately six out of ten analysed interactions expressed agreement, while disagreement represented slightly more than one third of responses. Agreement was particularly concentrated in posts related to political manipulation, criticism of media credibility, and institutional distrust. By contrast, posts focused on technological transformation and X as a communicative alternative displayed comparatively weaker levels of alignment and greater disagreement. This difference introduces a distinction between the two components identified above: agreement was more frequent around Musk's criticism of institutional media than around posts presenting X as a communicative alternative. The reception of anti-media claims and the reception of the platform proposed as an alternative therefore did not follow the same pattern.

At the same time, disagreement remained consistently visible throughout the interaction space. The distribution of responses suggests that interactions surrounding Musk's posts cannot be characterised exclusively as homogeneous spaces of reinforcement. Alongside support, the reply environment also contained substantial levels of contestation and disagreement.

The inferential analysis further suggests that users' positioning was only weakly associated with sociodemographic variables (RQ5). Although thematic orientation and professional background showed statistically significant relationships with agreement patterns, effect sizes remained limited. These findings suggest that variation in user positioning cannot be explained solely through demographic differentiation.

Several limitations should nevertheless be acknowledged. First, the study examines a period marked by intense electoral polarisation surrounding the 2024 United States presidential election and presidential transition. Second, demographic and occupational categories were inferred from publicly available profile information and should therefore be interpreted as approximate indicators rather than verified classifications. Given the proportion of profiles for which these variables could not be inferred, the observed associations apply to the classifiable profiles rather than to the broader population of X users. Finally, although AI-assisted classification enabled large-scale analysis, automated coding necessarily introduces interpretative limitations associated with ambiguity, irony, and context-dependent communication practices.

Furthermore, while the current frequency and inferential difference analyses were methodologically sufficient to evaluate semantic prevalence and thematic variations, the study's scope was confined to content-level discourse rather than relational network topology. Consequently, combining computational discourse analysis with Social Network Analysis (SNA) represents a fruitful avenue for future research to explore cluster formations, interaction graphs, and community-level dynamics across digital platforms.

Taken together, the findings contribute empirical evidence to ongoing discussions concerning anti-media discourse and changing forms of informational authority in digital environments. Rather than appearing as isolated media criticism, Musk's communication on X displays a recurrent discursive pattern in which criticism of journalism coexists with the positive construction of platform-based communication. This pattern can be situated within what Carlson et al. (2021) describe as competing claims to informational authority. Within the period analysed, these claims emerged not only through references to media organisations themselves, but also through competing representations of who should occupy a central role in the circulation and validation of information.

## 5 Conclusion

Musk's discourse constructs institutional journalism and X as competing sources of informational authority. Journalism is predominantly associated with bias, manipulation, and distrust, whereas X is presented through participation, direct communication, and user intervention. Citizen journalism thus functions not only as an appeal to wider participation, but also as a challenge to professional journalism's authority to mediate public information.

The resulting opposition, however, is not between mediation and direct communication. X remains an algorithmically organised and privately owned infrastructure that selects, orders, and distributes information. What changes is therefore the locus and logic of mediation: professional and editorial forms of selection are challenged in favour of platform-based mechanisms operating within an infrastructure controlled by the same actor who promotes it as an alternative.

User responses place a further limit on this displacement. Agreement was stronger around the criticism of institutional media than around the construction of X as its alternative, while disagreement remained substantial. "Cutting out the middleman" therefore captures the rhetoric more accurately than the process it describes: mediation is not removed but relocated, from professional and editorial gatekeeping towards a platform-based model that foregrounds direct user participation, even as ownership of the platform carries the capacity to shape the algorithmic mechanisms through which information gains visibility.

## Data availability statement

The raw data supporting the conclusions of this article will be made available by the authors, without undue reservation.

## Author contributions

UP: Writing – review & editing, Writing – original draft, Data curation, Validation, Supervision, Investigation. KM: Supervision, Software, Conceptualization, Writing – original draft, Project administration, Resources. SP: Resources, Writing – review & editing, Funding acquisition, Writing – original draft, Supervision. JM: Data curation, Methodology, Software, Writing – original draft, Visualization.

## Funding

The author(s) declared that financial support was received for this work and/or its publication. This research was funded by University of the Basque Country (EHU) through the project "Social Media, Disinformation and AI: Strategies and tools to prevent polarization and foster democratic values".

## Conflict of interest

The author(s) declared that this work was conducted in the absence of any commercial or financial relationships that could be construed as a potential conflict of interest.

## Generative AI statement

The author(s) declared that generative AI was not used in the creation of this manuscript.

Any alternative text (alt text) provided alongside figures in this article has been generated by Frontiers with the support of artificial intelligence and reasonable efforts have been made to ensure accuracy, including review by the authors wherever possible. If you identify any issues, please contact us.

## Publisher's note

All claims expressed in this article are solely those of the authors and do not necessarily represent those of their affiliated organizations, or those of the publisher, the editors and the reviewers. Any product that may be evaluated in this article, or claim that may be made by its manufacturer, is not guaranteed or endorsed by the publisher.